\documentclass[linenumbers,5p,twocolumn,authoryear]{elsarticle}
\usepackage{showyourwork}
\usepackage[version=4]{mhchem}
\usepackage{listings}
\usepackage{hyperref}
\usepackage{natbib}
\usepackage{xcolor}
\usepackage{orcidlink}

\definecolor{codegreen}{rgb}{0,0.6,0}
\definecolor{codegray}{rgb}{0.5,0.5,0.5}
\definecolor{codepurple}{rgb}{0.58,0,0.82}
\definecolor{backcolour}{rgb}{0.95,0.95,0.92}

\newcommand{\teff}{$T_{\rm eff}$}
\definecolor{tedcommentcolor}{HTML}{e17701}

\newcommand{\vspec}[1]{\texttt{VSPEC}#1}

\newcommand{\urldocsvspec}{\url{https://vspec-collab.github.io/VSPEC/}}
\newcommand{\urldocsvspecvsm}{\url{https://vspec-vsm.readthedocs.io/en/latest/}}
\newcommand{\urldocslibpypsg}{\url{https://tedjohnson12.github.io/libpypsg/}}

\begin{document}

% Title
\title{The \vspec{} Collection: A suite of utilities to model spectroscopic phase curves of 3D exoplanet atmospheres in the presence of stellar variability.}

\author[1,2,3,4,5]{Ted M Johnson\,\orcidlink{0000-0002-1570-2203}\corref{cor1}}
\ead{ted.johnson@unlv.edu}

\author[1,2,3]{Cameron Kelahan\,\orcidlink{0009-0002-0075-8041}}

\author[3]{Avi M. Mandell\,\orcidlink{0000-0002-8119-3355}}

\author[1,2,3]{Ashraf Dhahbi\,\orcidlink{0009-0009-0200-6821}}

\author[1,2,3]{Tobi Hammond\,\orcidlink{0009-0002-5756-9778}}

\author[3,6]{Thomas Barclay\,\orcidlink{0000-0001-7139-2724}}

\author[3]{Veselin B. Kostov\,\orcidlink{0000-0001-9786-1031}}
\author[3]{Geronimo L. Villanueva\,\orcidlink{0000-0002-2662-5776}}

\cortext[cor1]{Corresponding Author}

\affiliation[1]{
    organization={Southeastern Universities Research Association},
    addressline={1201 New York Avenue NW, Suite 430},
    city={Washington, DC},
    postcode={20055},
    country={USA}
}
\affiliation[2]{
    organization={Center for Research and Exploration in Space Science and Technology, NASA Goddard Space Flight Center},
    addressline={8800 Greenbelt Road},
    city={Greenbelt, MD},
    postcode={20771},
    country={USA}
}
\affiliation[3]{
    organization={NASA Goddard Space Flight Center},
    addressline={8800 Greenbelt Road},
    city={Greenbelt, MD},
    postcode={20771},
    country={USA}
}
\affiliation[4]{
    organization={Nevada Center for Astrophysics, University of Nevada, Las Vegas},
    addressline={4505 South Maryland Parkway},
    city={Las Vegas, NV},
    postcode={89154},
    country={USA}
}
\affiliation[5]{
    organization={Department of Physics and Astronomy, University of Nevada, Las Vegas},
    addressline={4505 South Maryland Parkway},
    city={Las Vegas, NV},
    postcode={89154},
    country={USA}
} 
\affiliation[6]{
    organization={University of Maryland, Baltimore County},
    addressline={1000 Hilltop Cir},
    city={Baltimore, MD},
    postcode={21250},
    country={USA}
}

\begin{abstract}
    We present the Variable Star PhasE Curve (\texttt{VSPEC}) Collection, a set of Python packages for simulating combined-light spectroscopic observations of 3-dimensional exoplanet atmospheres in the presence of stellar variability and inhomogeneity. \texttt{VSPEC} uses the Planetary Spectrum Generator's Global Emission Spectra (PSG/GlobES) application along with a custom-built multi-component time-variable stellar model based on a user-defined grid of stellar photosphere models to produce spectroscopic light curves of the planet-host system. \texttt{VSPEC} can be a useful tool for modeling observations of exoplanets in transiting geometries (primary transit, secondary eclipse) as well as orbital phase curve measurements, and is built in a modular and flexible configuration for easy adaptability to new stellar and planetary model inputs. We additionally present a set of codes developed alongside the core \texttt{VSPEC} modules, including the stellar surface model generator \texttt{vspec-vsm}, the stellar spectral grid interpolation code GridPolator, and a Python interface for PSG, \texttt{libpypsg}.
\end{abstract}


% Highlights:
% Instruments that characterize exoplanet atmospheres are sensitive to the effects of stellar inhomogeneities.
% We present VSPEC, a code to simulate observations of 3D planets in the presence of stellar variability and inhomogeneity.
% VSPEC, and the codes that support it, are open source, well documented, and available to the public.

\maketitle

\section{Introduction}
\label{sec:intro}

In the era of high-sensitivity transiting exoplanet characterization missions such as the Hubble Space Telescope (HST), the James Webb Space Telescope (JWST), and the future Atmospheric Remote-sensing Infrared Exoplanet Large-survey (ARIEL), spectral analysis of exoplanet atmospheres is increasingly sensitive to contamination due to stellar inhomogeneities (e.g. spots, granulation) and stochastic stellar variability (e.g. flares).  As an example, recent analysis of the JWST/NIRSpec transit of GJ 486b by \citet{moran2023} exposed a degeneracy between atmospheric absorption by water and water-rich spots on the stellar surface, and similar effects were seen in the transit spectrum of LHS 1140b \citep{cadieux2024} and many other recent transit measurements with JWST \citep[e.g.,][]{lim2023,may2023,fournier-tondreau2024}. This ``transit light source effect'' \citep[TLS,][see also \citet{apai2018,barclay2021,garcia2022,barclay2023}]{rackham2018} occurs when the region of the stellar surface occulted by the transiting planet is not representative of the disk-integrated spectrum.

Stellar contamination may also cause errors in the extraction of thermal phase curve spectroscopy if the stellar spectrum changes significantly over the period of the planet's orbit; if the variations are non-linear, but only linear interpolation between subsequent secondary eclipses is used to remove the stellar contribution, the resulting planetary flux measurements will not be exclusively representative of the planet's emission. This will be particularly important for observations of planets around more variable stars such as active M-stars or later-type evolved host stars. %the Mid-IR Exoplanet CLimate Explorer \citep[MIRECLE,][]{mandell2022} mission concept will employ the Planetary Infrared Excess \citep[PIE][]{stevenson2020} technique to extract the planetary contribution from combined-light observations. Uncertainties on the stellar spectrum will dominate the analysis if it is not removed appropriately.

To adequately prepare for future observations and future missions, it will be necessary to demonstrate methods to mitigate these effects. This task requires a flexible tool that combines models of exoplanet atmospheres and stellar variability in a robust way. In this paper we present the \vspec{} (Variable Star PhasE Curve\footnote{\url{https://github.com/orgs/VSPEC-collab/repositories}}) Collection, a collection of open-source Python 3 packages to simulate observations of exoplanet systems with variable host stars. The \vspec{}\footnote{\url{https://github.com/VSPEC-collab/VSPEC}} package itself is merely an interface that pulls together a variable star surface model \texttt{vspec-vsm}\footnote{\url{https://github.com/VSPEC-collab/vspec-vsm}}, a stellar spectra grid interpolation code Gridpolator\footnote{\url{https://github.com/VSPEC-collab/GridPolator}}, and a Python interface for the Planetary Spectrum Generator \citep[PSG,][]{villanueva2018}, \texttt{libpypsg}\footnote{\url{https://github.com/tedjohnson12/libpypsg}}. All of these codes together allow \vspec{} to account for 3D planetary atmospheres, time-resolved effects of both planet and star, multiple geometries including transit and secondary eclipse, and realistic noise modeling.

In this paper we will describe the science motivation behind all of these codes, the details of each package and module in the \vspec{} collection, and demonstrate examples of their use. Section \ref{sec:vspec} describes \vspec{} and goes into some detail about setup, use, and the code's internals. In Section \ref{sec:libpypsg} we describe \texttt{libpypsg} as an interface for the Planetary Spectrum Generator. Section \ref{sec:star} describes \vspec{}'s stellar model, \texttt{vspec-vsm}. We discuss the spectral model interpolation code GridPolator in Section \ref{sec:gridpolator}. Section \ref{sec:vspec-gcm} is dedicated to the simple climate model that \vspec{} uses to produce realistic phase curves in the absence of a user-supplied output file from a more sophisticated Global Circulation Model (GCM). Section \ref{sec:examples} demonstrates a few simple \vspec{} use cases. Finally, in Section \ref{sec:conclusion} we will discuss the future of the code and issues it might have.

\section{Using \vspec{}}
\label{sec:vspec}

\begin{figure*}
    \includegraphics[width=\textwidth]{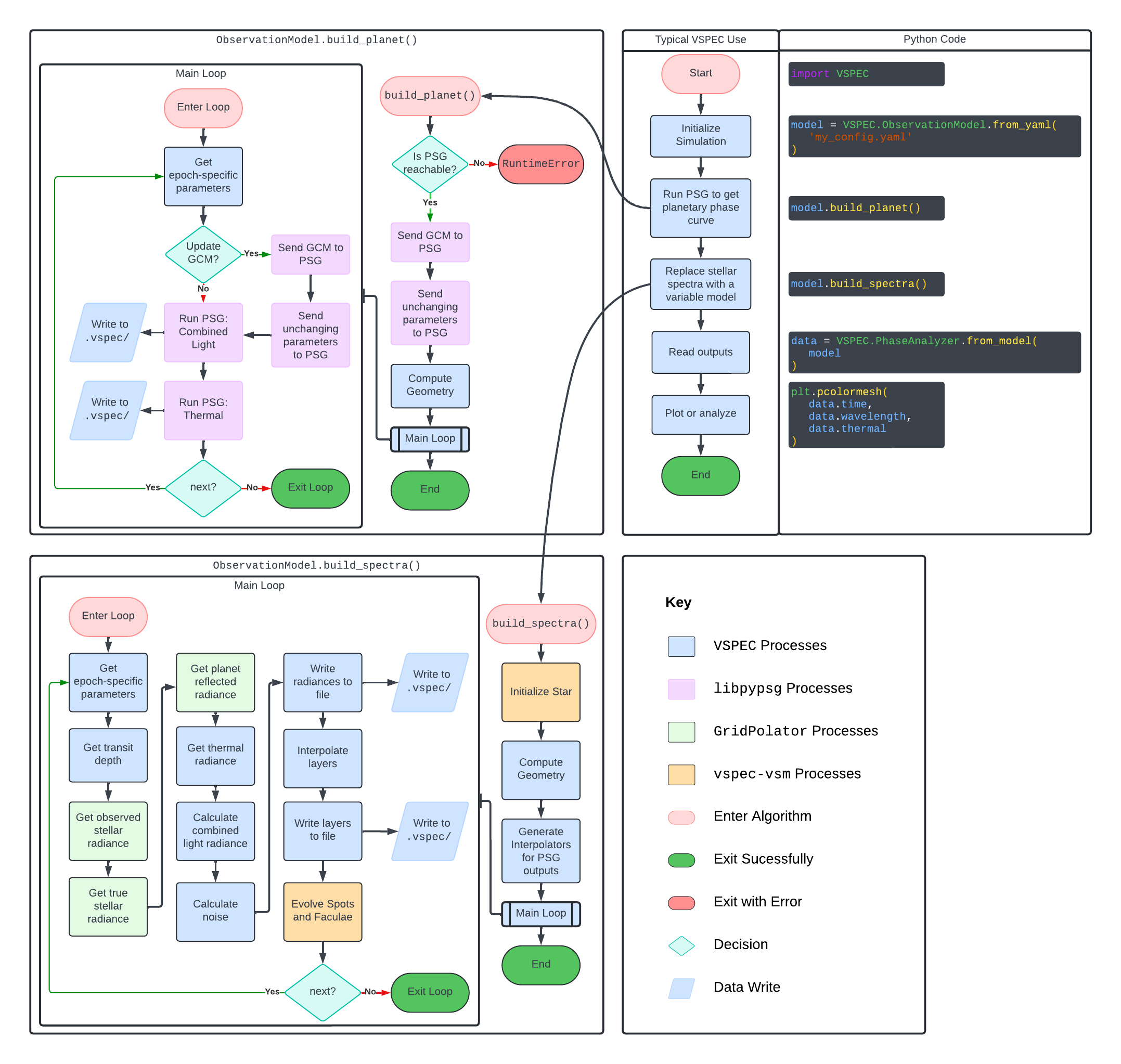}
    \caption{
        Flowchart diagram outlining the typical workflow of a \vspec{} simulation.
        {\bf Upper right}: Processes representing the main \vspec{} stages alongside Python code snippets. Nearly all \vspec{} use cases involve the code \texttt{model.build\_planet(); model.build\_spectra()}, though the input file and the particulars of the analysis can be very different.
        {\bf Upper left}: \texttt{build\_planet()} algorithm. This portion of the code is responsible for interfacing with PSG (via \texttt{libpypsg}, see Section \ref{sec:libpypsg}) to generate a phase curve. The decision labeled ``Update GCM?'' evaluates to true when a time-dependent climate model is specified in the input file, or when the PSG configuration returned by the latest API call is longer than 1500 lines, indicating that PSG should be reset. For convenience, the PSG GlobES application appends new configuration files to those from previous API calls, but stops reading after 2000 lines for security reasons. Therefore, the internal configuration file must be reset on occasion.
        {\bf Lower left}: \texttt{build\_spectra()} algorithm. This portion of the code is responsible for adding stellar variability into our results. We use the outputs from the \texttt{build\_planet()} step to create interpolator instances (implemented as \texttt{GridPolator.Gridspectra} objects, see Section \ref{sec:gridpolator}) for the time- and wavelength-dependent values of relevant quantities (e.g., thermal emission flux, transit depth, various types of noise). We then step through each epoch in our time series and evaluate the variable star model (using \texttt{vspec-vsm}, see Section \ref{sec:star}). We replace the stellar spectrum from PSG with that of our model, scaling the noise appropriately. Finally, we evolve the star to the time of the next observation.
        {\bf Lower right}: Diagram key. Note that, in general, all the steps marked as \texttt{GridPolator} processes also involve \texttt{vspec-vsm}.
    }
    \label{fig:vspec-diagram}
    \script{copy_static_figs.py}
\end{figure*}

The typical flow of \vspec{}'s code is shown in Figure \ref{fig:vspec-diagram}, as well as described in the following sections.
It is generally broken into four parts: i) read in the configuration, ii) compute planetary spectra, iii) compute stellar spectra, iv) analyze, with steps (i) and (iv) varying widely by use case.

Most users of \vspec{} will interact exclusively with two classes: \texttt{VSPEC.ObservationModel} and \texttt{VSPEC.PhaseAnalyzer}. The bulk of the work is done by the \texttt{ObservationModel} class, as it reads in a user's inputs, runs the model, and writes output files. \vspec{} produces many output files, which themselves can be read easily using tools such as Astropy \citep{astropycollaboration2013,astropycollaboration2018,astropycollaboration2022}. However, the \texttt{PhaseAnalyzer} class makes it easy to read them as an ensemble, as well as quickly pull out portions of the data that are of interest.

Section \ref{subsec:config} describes initializing the \texttt{ObservationModel}. Section \ref{subsec:phase-curve} describes computing the planetary phase curve using the \texttt{ObservationModel.build\_planet()} method. The next step, calling \texttt{ObservationModel.build\_spectra()}, recalls the data produced by PSG previously, but combines it with a spectral model of a variable star, scaling reflected light and noise appropriately. This step, along with detailed descriptions of various substeps, is described in Sections \ref{subsec:stellar-lc}, \ref{subsec:reflected}, \ref{subsec:noise}, and \ref{subsec:transit}.

Use of the \texttt{PhaseAnalyzer} object is not described in detail as it mostly acts as a container, but Section \ref{sec:examples} includes many examples of its use. Additionally, the \vspec{} documentation page\footnote{\urldocsvspec} contains detailed API documentation.
\subsection{Configuring \vspec{}}
\label{subsec:config}

\vspec{} configurations are designed to minimize human errors by providing two equivalent formats. The first, a file written in YAML, is optimized for
human readability. For example, the \texttt{system} section, which describes the relationship between the observed planetary system and the observer, could
be written as shown in Listing \ref{ls:yaml}.
\begin{lstlisting}[label={ls:yaml},caption=\vspec{} YAML Configuration]
system:
    distance: 12.4 pc
    inclination: 89.7 deg
    phase_of_periastron: 0 deg
\end{lstlisting}
Note that the units of each of these parameters is included in a human-readable way. Internally, \vspec{} casts each of these parameters to an Astropy
\texttt{Quantity} instance, so that the user can input any desired value and unit combination so long as the physical type is correct -- i.e. it would be
equivalent to write \texttt{3.83e17 m} as the value for the \texttt{distance} parameter.

The second input method available is to directly initialize the Python object that \vspec{} uses internally to store its model parameters.
The top-level object, \texttt{InternalParameters}, is structured to mirror the YAML input file, with one argument per YAML section. The same
configuration shown in Listing \ref{ls:yaml} would be written in Python as shown in Listing \ref{ls:python}.
\begin{lstlisting}[language=Python, caption={\vspec{} Python configuration},label={ls:python}]
from astropy import units as u
from VSPEC.params import *
params = InternalParameters(
    ... # Other arguments
    system=SystemParameters(
        distance=12.4*u.pc,
        inclination=89.7*u.deg,
        phase_of_periastron=0*u.deg
    )
)
\end{lstlisting}
This is convenient for producing configurations programmatically, for example for producing a grid of model phase curves.

However the user decides to provide model parameters, they are read into the main \vspec{} object: the \texttt{ObservationModel}.
Upon initialization a local working directory (by convention \texttt{.vspec/}) is created to store simulation outputs and intermediate files.

\subsection{Planetary Phase Curve}
\label{subsec:phase-curve}
\vspec{} generates a phase curve when the \texttt{ObservationModel.build\_planet()} method is called. It does this though a series of API calls to the Planetary Spectrum Generator \citep[PSG][]{villanueva2018}, using \texttt{libpypsg} (see Section \ref{sec:libpypsg}) to interface with the API. Initially, configurations are sent that give PSG information about the system that does not change with time -- that includes the 3D Global Circulation Model (GCM) and instrument parameters. \vspec{} then enters the main loop, iterating through each observation epoch (e.g.integration or combination of integrations) and making a pair of API calls: one to get a combined light spectrum and one to get the thermal flux only. Results from PSG in each epoch are stored locally in the \texttt{.fits} format. 

\subsubsection{Phase Sampling}
Making API calls to PSG can be computationally expensive, especially when trying to resolve small structures in the climate model. In order to reduce unnecessary calls to PSG, the temporal sampling of the planetary spectrum is independent of the stellar model/output sampling. When computing the planetary flux in the output files, the raw PSG output is interpolated and averaged over each time step using the trapezoid integration rule. By default one spectrum is computed with PSG at each interface separating time steps; this eases the calculation of edge effects between steps.

\subsection{Stellar light curve}
\label{subsec:stellar-lc}

The last step is to replace the stellar spectra used by PSG with \vspec{}'s stellar model by calling the \texttt{ObservationModel.build\_spectra()} method. In this step \vspec{} acts as a bridge between
the stellar surface model \texttt{vspec-vsm} (which models spatial changes to the star with time, see Section \ref{sec:star}) and a grid of
pre-computed stellar spectra (from the GridPolator package; see Section \ref{sec:gridpolator}). The star is initialized and allowed to evolve for a user-specified amount of time in order for spots and faculae to approach growth-decay equilibrium (this is also called the ``burn in'' phase). The star is then evolved at the output cadence, generating a set of key-value pairs at each epoch that describe the surface temperature and coverage fractions of the portion of the stellar disk visible to the observer. These \teff s and coverage fractions are fed into GridPolator to produce composite stellar spectra.

\subsection{Reflected light}
\label{subsec:reflected}
\vspec{} then produces a composite spectrum of the portion of the star visible to the planet to accurately incorporate the time-dependent stellar flux into the reflected-light spectrum. To compute the total reflected light in the output files, \vspec{} first divides the reflected light flux from PSG by the PSG stellar model to obtain the apparent albedo (I/F). This albedo is then multiplied by the composite stellar spectrum (facing the planet) to obtain the total reflected flux.

\subsection{Simulated Noise}
\label{subsec:noise}
Alongside the planetary radiance files, PSG returns a file giving the analytic noise estimate for each epoch broken down by source. We assume that the total noise is the quadrature sum of its constituent parts and that the \texttt{source} column (i.e. photon noise) is the only one affected by replacing one stellar spectrum with the other. This means the photon noise can be calculated:
\begin{equation}
    N_{\rm photon} = N_{\rm photon,~PSG} \sqrt{\frac{F_{*,\rm ~VSPEC}}{F_{*,\rm ~PSG}}}
\end{equation}
where $F_*$ is the stellar flux. The total noise associated with the integration is then
\begin{equation}
    N_{\rm tot} = \sqrt{
        N_{\rm photon}^2 + N_{\rm detector}^2 + N_{\rm telescope}^2 + N_{\rm background}
    }
\end{equation}

\subsection{Transit and Eclipse}
\label{subsec:transit}
Users must be careful when using \vspec{} to simulate transit and eclipse measurements because of the vast timescale differences between those events and the typical phase variations of a planet -- a sparsely sampled phase curve that happens to include one epoch of transit will not adequately sample the light curve accurately during a transit or eclipse because the interpolator knows no better than to assume a linear interpolation scheme. It is recommended that users use a high cadence for observations of transits and eclipses so that each event will be sampled by more than one point.

In the case of a transiting geometry, \vspec{} computes the spectrum of the occulted portion of the star in order to properly simulate the transit light source effect \citep[TLS][]{rackham2018}. This spectrum is the flux blocked by the planet in the case that it is a solid occulting circle, and therefore produces a flat transmission spectrum. The effect of atmospheric transmission is added by comparing the PSG-computed effective radius to a purely geometric calculation of such a ``bare-rock" transit depth:
\begin{equation}
    F_{\lambda} = F_{{\rm rock}, \lambda}\, F_{{\rm PSG,} \lambda} \,\left (\frac{R_*}{R_p} \right )^2
\end{equation}
where $F_\lambda$ is the flux of the variable star surface occulted by the planet with the atmosphere considered. $F_{{\rm rock}, \lambda}$ is the flux of the surface occulted by a bare rock with the planet's radius, $F_{{\rm PSG,} \lambda}$ is the flux computed by PSG to be lost
due to occultation and transmission, and $R_*$ and $R_p$ are the stellar and planetary radii, respectively.

In the case of a total eclipse of the planet, the planetary thermal and reflected contributions are set to zero flux. However, in the case of a partial eclipse (large impact parameter), \vspec{} computes the fraction of the planet that is visible to the observer and reduces both thermal and reflected flux accordingly. In this case \vspec{} treats the planet homogeneously; however, future work could utilize PSG/GlobES's ability to return a hypercube of spatially resolved spectra in order to construct planetary spectra only from regions of the planet that are not eclipsed.

\section{Interfacing with the Planetary Spectrum Generator via \texttt{libpypsg}}
\label{sec:libpypsg}
PSG\footnote{\url{https://psg.gsfc.nasa.gov/}} \citep{villanueva2018} is a powerful radiative transfer tool that is ubiquitous in the exoplanet and solar system atmosphere fields. Either through its web graphical interface, its public API, or the local version available through Docker, PSG allows users to simulate observations of planets, comets, and moons with a variety of geometries and realistic noise models.

\begin{lstlisting}[
    language=Python,label={ls:libpypsg},
    caption={\texttt{libpypsg} Example Input}
]
import libpypsg
from astropy import units as u
from netCDF4 import Dataset

# This next line only needs to be run
#     once per user.
#     These settings are 
#     saved to `~/.libpypsg/settings.json`
libpypsg.settings.save_setting(
    api_key='my_api_key',
    url='https://psg.gsfc.nasa.gov/api.php'
)

gcm_data = Dataset('path/to/gcm.nc')

cfg = libpypsg.cfg.PyConfig(
    target=libpypsg.cfg.Target(
        object='Exoplanet',
        name='proxima-cen-b'
        ...
    ),
    geometry=libpypsg.cfg.Observatory(...),
    generator=libpypsg.cfg.Generator(...),
    telescope=libpypsg.cfg.SingleTelescope(...)
    noise=libpypsg.cfg.CCD(...)
    gcm=libpypsg.globes.waccm_to_pygcm(
        gcm_data, itime=0,
        molecules=['H2O','CO2'],
        aerosols=['Water','WaterIce'],
        mean_molecular_mass=48.0)
)

caller = libpypsg.APICall(
    cfg=cfg, app='globes',
    output_type='rad'
)
response = caller()
print(response.rad['Wave/freq','proxima-cen-b'])
\end{lstlisting}

\vspec{} uses a Python interface for the PSG API system called \texttt{libpypsg}, which is a general stand-alone library built for any Python-based API call to PSG. \texttt{libpypsg} draws inspiration from object-relational mapping in frameworks such as Django\footnote{\url{https://www.djangoproject.com/}} to encode a PSG configuration file as a native Python object. Fields of a \texttt{libpypsg} data model represent one or more lines of a configuration file, and the two representations are interchangable without loss of information. Upon creation of this configuration as a \texttt{PyConfig} object, a user can create an \texttt{APICall} instance which handles sending a request to PSG via Python's \texttt{requests}\footnote{\url{https://github.com/psf/requests}} package. This dedicated caller is useful because it reads a user's settings (e.g. an API key or the URL of the PSG API endpoint) from a file on their hard drive. This makes it much simpler to publish code that utilizes PSG because 1) it eliminates the danger of accidentally publishing an API key, and 2) it means that users can run the same code regardless of whether they have a local installation of PSG.

\begin{lstlisting}[caption={\texttt{libpypsg} Example Output},label={ls:psgout}]
Wave/freq  proxima-cen-b
    um      W / (um m2) 
----------- -------------
        1.0   1.81314e-20
       1.02    3.1926e-20
     1.0404   4.95983e-19
    ...           ...
16.97557767   7.54221e-20
17.31508922   7.49258e-20
17.66139101   7.35507e-20
Length = 146 rows

\end{lstlisting}

A powerful feature of \texttt{libpypsg} is that it provides a universal interface for 3D atmosphere models and PSG's Global Emission Spectra (GlobES) application (\citealt{fauchez2024}, in prep; see also \citealt{kofman2024}, accepted)
GlobES expects 3D data to be presented in a binary data format, and converting a model to this format is non-trivial. \texttt{libpypsg}'s \texttt{PyGCM} class provides a native Python representation of GlobES input, and includes built-in methods for converting popular climate models like ExoCAM \citep{wolf2022}, the Whole Atmosphere Community Climate Model \citep[WACCM,][]{marsh2013}, and ExoPlaSim \citep{paradise2022} to the correct format.

Listing \ref{ls:libpypsg} shows a minimal example that uses \texttt{libpypsg} to make a call to PSG/GlobES, with the output shown in Listing \ref{ls:psgout}. Here \texttt{libpypsg} converts a GCM output from the netCDF format to a \texttt{PyGCM}, which, along with data model classes like \texttt{Target} and \texttt{Observatory}, is used as an argument to construct a \texttt{PyConfig} class instance. This configuration object is then passed to an \texttt{APICall} instance. When the API call is made, the contents of the \texttt{PyConfig} are included in the body of the POST request in the \texttt{file} field. PSG reads this the same as it would if we had written those contents to a file \texttt{config.txt} and passed the \texttt{--data-urlencode file@config.txt} option to \texttt{cURL}, as are the instructions in the PSG handbook \citep[pp. 162-163]{villanueva2022}. More details on use of this library including its API and examples can be found on the \texttt{libpypsg} documentation page\footnote{\urldocslibpypsg}.

\section{Stellar Variability Model}
\label{sec:star}

The \vspec{} stellar model is designed in a modular fashion to allow for both simple and complex behaviors. The model is available as the Python package \texttt{vspec-vsm}, which is based on the earlier stellar spot model \texttt{spotty} \citep{barclay2021}. It is spectral-model agnostic, meaning that it relies on another package (GridPolator, see Section \ref{sec:gridpolator}) to compute the actual variable spectra. Instead, the surface model takes a pair of latitude and longitude points as input and returns a dictionary $\{ T_{\rm eff} : f_{T_{\rm eff}} \}$ where the keys \teff~ are the effective temperatures of each surface component and the values $f_{T_{\rm eff}}$ are the fraction of the visible stellar disk with that effective temperature.

The main model class is the \texttt{Star} object, which describes bulk properties (e.g. radius, photospheric \teff, limb darkening parameters) in addition to acting as 
a container for objects representing sources of variability such as \texttt{StarSpot}, \texttt{Facula}, \texttt{StelarFlare}, and \text{Granulation}, which are described below. For a complete description of the code, including examples, see the \texttt{vspec-vsm} documentation page\footnote{\urldocsvspecvsm}.

\subsection{Stellar Surface Grid}
Sources of variability that exist as major structures on the stellar surface (i.e. spots and faculae) are encoded as deviations from the photospheric effective temperature on a grid of points representing the stellar surface. Each time the fractions of each surface component are calculated, the grid is initialized with constant \teff. The code then iterates through each spot and facula, replacing portions of the grid with the effective temperatures that make up each structure.

The simplest grid to build is one that is rectangular in latitude and longitude, and in which the area that each point represents decreases towards the poles. However, this can lead to unnecessary spatial oversampling at the poles, and may lead to poor resolution at the equator to keep the total number of points from being excessively high. To mitigate these problems, \texttt{vspec-vsm} supports a ``grid'' generated using a Fibbonacci spiral to create a near-isotropic distribution of points. This is especially useful when simulating transits, giving optimal sampling at the stellar equator. It is also useful for resolving small structures on the surface, as the number of points needed for a given minimum resolution is greatly reduced.

\begin{figure}
    \centering
        \includegraphics[width=0.5\textwidth]{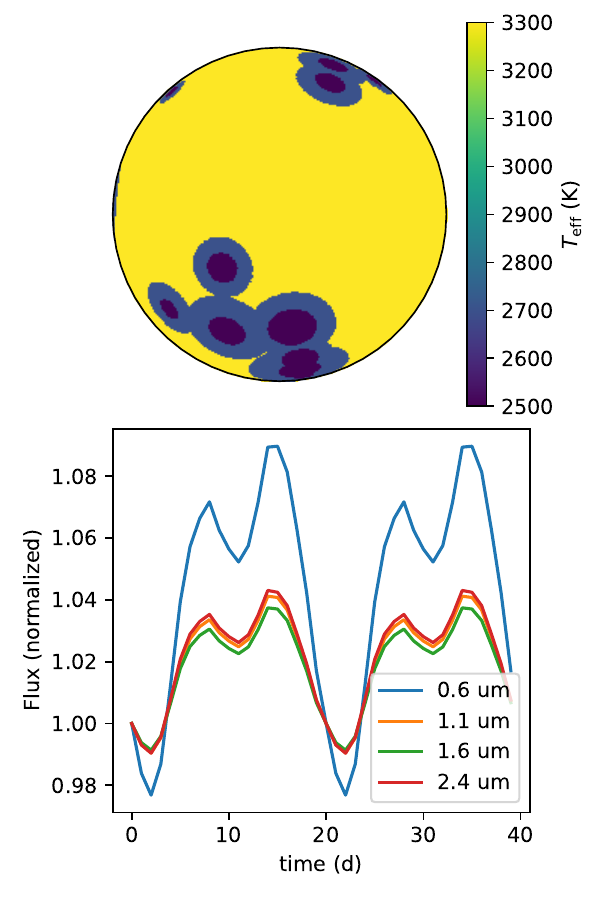}
    \script{surface_map_and_lc.py}
    \caption{{\bf Top}: The surface map of a star with spots. The star has a photospheric temperature of 3300K, a penumbra temperature of 2700K, and an umbra temperature of 2500K.
    {\bf Bottom}: An example light curve of a spotted star. Flux variations are due to varying spot coverage as a function of sub-observer point. Two rotational periods are shown.
    }
    \label{fig:surface_map}
\end{figure}

\subsection{Spots \label{subsec:spots}}
Our star spot model is nearly entirely based on observations of the Sun. Sunspots can be resolved and are well-studied, whereas spots on other stars (especially non-solar-type stars) can only be observed indirectly. We therefore designed our spot model to mimic the behavior of sunspots but with parameterized values for spot temperature and lifetime that can be matched to observations of other stellar types and ages. On the Sun, spots have two regions shown in Figure \ref{fig:surface_map}: the dark, central umbra and the lighter, surrounding penumbra (a detailed review of sunspot behavior, including sizes and lifetimes, is described in \citet{solanki2003}). The spot area as a function of time is
\begin{equation}
    A(t) = \left\{
    \begin{array}{lr}
        A_0 \exp{((t-t_0)/\tau)}, & \text{if } t \leq t_0 \\
        A_0 - W(t-t_0), & \text{if } t > t_0
    \end{array}
    \right\}
\end{equation}
where $A_0$ is the maximum area reached, $t_0$ is the time of the maximum, $\tau$ is the exponential growth rate, and $W$ is the linear decay rate.

In addition to modeling the variability produced by a population of spots, \texttt{vspec-vsm} contains utilities to produce those populations given some description.
These parameters include the average spot area \citep[lognormally distributed, ][]{bogdan1988}, \teff, growth and decay rates, and the method for distributing spots on the surface. These \texttt{SpotGenerator} objects are responsible for evolving the spot population as the star ages. The number of spots created in some interval $\Delta t$ is:
\begin{equation}
    N(\Delta t) = \frac{4 \pi R_*^2 f_{\rm spot}}{A_{\rm mean}} \frac{\Delta t}{\tau_\text{spot}}
\end{equation}
where $R_*$ is the stellar radius, $f_{\rm spot}$ is the fraction of the surface covered by spots in growth-decay equilibrium, $A_{\rm mean}$ is the lifetime-averaged mean spot area, and $\tau_\text{spot}$ is the spot lifetime, defined as:
\begin{equation}
    \tau_\text{spot} = \frac{A_0}{W} - \frac{1}{\tau} \ln{\left (\frac{A_{\rm min}}{A_0}\right )}
\end{equation}
where $A_{\rm min}$ is the area given to each spot when it is initialized. There are also two modes to distribute spots on the star's surface. The first distributes them isotropically; the second, modeled after analysis of spots on the Sun by \citet{mandal2017}, concentrates the spot latitude around $\pm 15^{\circ}$.

\subsection{Faculae \label{subsec:faculae}}
\begin{figure}
    \centering
    \includegraphics[width=0.5\textwidth]{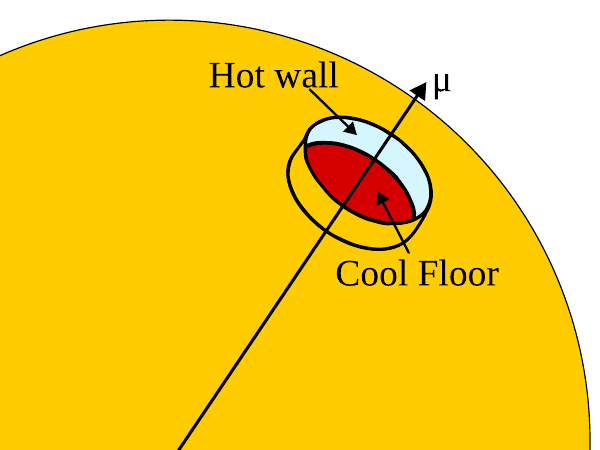}
    \includegraphics[width=0.5\textwidth]{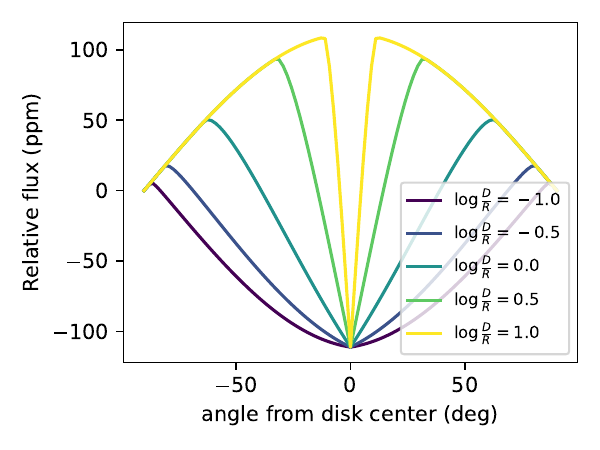}
    \script{facula_depth.py}
    \caption{
        ``Hot Wall'' model of faculae. Faculae structure causes their contrast to be dependent on their distance from the center of the disk. {\bf Top}: Depiction of a facula on the limb of a star. The hot wall is exposed to the observer causing the pore to appear bright. At disk center, the cool floor is most visible. {\bf Bottom}: The effects of depression depth and viewing angle on facula brightness. The 3D structure of faculae is most apparent when radius $\sim$ depth. A toy flux model was used to demonstrate the shape of these curves, but in practice their magnitudes depend on stellar spectral models.
        }
    \label{fig:fac_struct}
\end{figure}

Faculae are magnetically-generated regions of the solar surface that usually appear as bright points near the limb; we employ the ``hot wall''
model \citep{spruit1976} where faculae are described as three-dimensional pores in the stellar surface with a hot, bright wall and a cool, dark floor, as shown in Figure \ref{fig:fac_struct}. Their three-dimensional structure causes faculae's observational properties to change depending on their angle from disk-center. Close to the limb,
the hot wall is visible to the observer, and faculae appear as bright points; near the center, however, the cool floor is exposed and faculae appear dark. To consider this effect in the faculae light curve, we compute the fraction of each facula's projected area -- the area on the disk it would occupy as seen by the observer -- that is occupied by the hot wall versus the cool floor. Mathematically, this is no different from finding the area of the intersection of two overlapping ellipses, and we use this conceptualization to calculate the contributions of each portion, as described below.

As shown in Figure \ref{fig:fac_struct}, peering through the top of the pore the observer can see a portion of the hot wall and a portion of the cool floor. Consider an ellipse centered at the origin with semimajor axis $R$ and semiminor axis $\mu R$, oriented so that the semimajor axis points along the $y$-axis. Now consider a second, identical ellipse, displaced along the $x$-axis by some amount $0<d<\mu R$. The intersection of these two ellipses is the portion of the cool floor that is visible, and the area of the second ellipse outside the intersection is the portion of the hot wall that the observer sees. We are not interested in the absolute areas of each of these regions, only how they relate to the total ellipse area $\pi \mu R^2$, and so we can generalize to the case that $\mu=1$ (i.e. circles). This problem is solved for arbitrary circles by \citet{weisstein2004}; the details of their derivation are not relevant to this text, but we use their result to create a function $f(d,R)$ that computes the fraction of a circle with radius $R$ and center $(d,0)$ that is inside the unit circle. For a facula with radius $r$, wall height $h$, and angle from disk center $\theta$, we find that the effective area of the cool floor is:
\begin{equation}
    \frac{A_{\rm floor}}{A_{\rm total}} = f\left(\frac{h}{r}\tan{\theta}, 1 \right)
\end{equation}
and for the hot wall it is:
\begin{equation}
    \frac{A_{\rm wall}}{A_{\rm total}} = 1-  f\left(\frac{h}{r}\tan{\theta}, 1 \right)
\end{equation}

According to studies of solar faculae \citep{topka1997}, faculae temperatures (of both the floor and wall) are dependent on the facula radius, while depth appears to be constant. They also find that the smallest faculae have no visible floor, and that even at disk center they appear as bright points. We parameterize the floor temperature to be
\begin{equation}
    \Delta T_{\rm eff, floor} = \left\{
    \begin{array}{lr}
    \text{Not visible}, & \text{if } r<r_{\rm min} \\
    m_{\rm floor}(r-r_{\rm min}) + \Delta T_{\rm eff, floor, 0}, & \text{if } r \ge r_{\rm min}
    \end{array}
    \right\}
\end{equation}
where $r$ is the radius of the facula, $r_{\rm min}$ is the minimum radius where the floor is visible, $\Delta  T_{\rm eff, floor, 0}$ is the difference between the floor and photosphere \teff at $r_{\rm min}$, and $m_{\rm floor}$ is the slope of the relationship with units of [temperature] [length]$^{-1}$. Similarly, the wall temperature is parameterized as
\begin{equation}
    \Delta T_{\rm eff,wall} = m_{\rm wall}r + \Delta T_{\rm eff,wall,0}
\end{equation}
where $\Delta T_{\rm eff,wall,0}$ is the temperature of a zero-radius facula and $m_{\rm wall}$ is the slope of the radius-temperature relationship with units of [temperature] [length]$^{-1}$. These relationships can be defined by the user, for example setting $m=0$ for constant temperatures.

Facula lifetimes are defined as the time it takes its radius to decay by $e^{-2}$. Because faculae grow and decay exponentially at the same rate, each facula spends one lifetime with a radius greater than $e^{-1}$ of its maximum. Each facula is born and dies at a radius of $e^{-2}$ of its maximum, effectively existing in the code for two lifetimes. \citet{hovis-afflerbach2022} suggest a typical facula lifetime to be on the order of 6 hours, with a distribution that resembles a Poisson function; however we choose a log-normal distribution for both lifetime and maximum radius because it does not allow for these values to be 0. We choose to correlate lifetime and radius so that they are determined by a normalized random distribution - for each new facula, a value is randomly drawn and determines both the facula lifetime and maximum radius.

\subsection{Flares \label{subsec:flares}}
Flares are an important source of stellar variability on short timescales. \texttt{vspec-vsm}'s flare light curve model is based on the \texttt{xoflares} package \citep{barclay2020}, which itself is based on empirically derived light curve shapes from \citet{davenport2016}. However, \texttt{xoflares} is designed to fit light curves in a single spectral band, so we add additional parameters to extend it to a multiwavelength light curve. We completely describe a flare using its temperature, total energy, full width at half maximum (FWHM), and the time of its peak. We model a flare as a hot, optically thin region above the photosphere that produces a blackbody spectrum. In this model, the temperature is constant, and the sharp rise and fall seen in the light curve is caused by the region's rapidly changing area. This simplification allows the total flare energy $E$ to be used as a normalization factor:
\begin{equation}
    \int_{-\infty}^{\infty}A\,dt = \frac{E}{\sigma T^4}
\end{equation}
where $A$ is the time-dependent area of the flare region, $T$ is the constant flare temperature, and $\sigma$ is the Stefan-Boltzmann constant. This allows flare light curves to be produced with a fixed bolometric luminosity. Because the temperature is fixed, the relevant quantity for modeling the flux of a flare in a given integration is the integrated time-area $\int A\,dt$. These quantities are computed numerically and fed to \vspec{} along with $T$ to produce spectra with the appropriate absolute flux.

\begin{figure}
    \centering
    \includegraphics[width=0.5\textwidth]{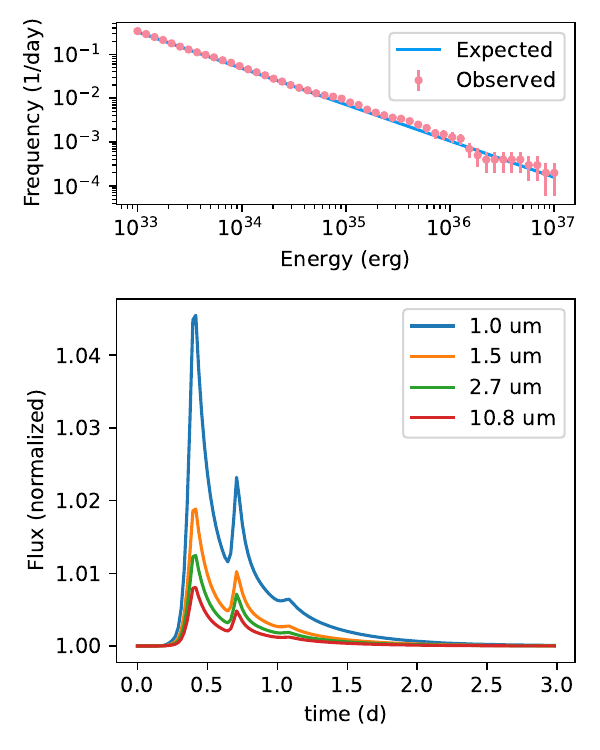}
    \caption{
        {\bf Top}: Generated flare frequencies compared to those expected from \citet{gao2022}, generated using a 10,000 day simulation. The $y$-axis shows the frequency of flares with energies greater than or equal to the value of the $x$-axis. Error bars based on the square root of the number of observed flares as this is a Poisson process. {\bf Bottom}: light curve of a star flaring with the same power-law slope as the top panel, but the intercept ($\beta$) has been increased by 0.3 for visual effect. The mean flare temperature is $9000$ K and the mean FWHM is $3$ hours.
        }
    \script{flare_freq.py}
    \label{fig:flare_freq}
\end{figure}

Before simulating an observation, \vspec{} asks \texttt{vspec-vsm} to pre-compute a population of flares based on a power-law frequency-energy relationship based on Kepler and TESS light curves \citep{gao2022}:
\begin{equation} \label{eq:flare_freq}
    \log{(f/{\rm [day]})} = \beta + \alpha~\log{(E/{\rm [erg]})}
\end{equation}
where $f$ is the frequency of flares with energies $\ge E$. We compute the number of expected flares over a time duration and determine the number to create $N$ by a random Poisson draw. We then generate $N$ flares with energies determined by the quantile function:
\begin{equation}
    E = E_{\text{min}} (1-X)^{1/\alpha}
\end{equation}
where $E_{\text{min}}$ is the minimum considered flare energy and $X$ is a random number on the interval $[0,1)$. We set the default values $\alpha=-0.829$, $\beta=26.87$ from \citet{gao2022}, but these can be adjusted by the user. Figure \ref{fig:flare_freq} shows our simulation results compared to their power-law values.

\subsection{Granulation}
Granulation is a source of stellar variability that arises from convection near the surface of the star. The result is a stellar surface that is not constant in temperature and that changes on very short timescales. Hydrodynamic simulations of stellar atmospheres \citep[e.g.][]{magic2014} describe hot granules of rising gas surrounded by cooler, sinking regions with a temperature a few percent lower than the nominal value. 

We model granulation as a global process, and its effects are computed after the effects of spots and faculae. Of the remaining ``quiet'' photosphere (i.e. the regions not covered in spots of faculae), a fraction is computed to be part of the cool
inter-granule surface; the surface coverage of the cool region at any given time is computed by a Gaussian process (GP) using the {\sc tinygp} package \citep{foreman-mackey2024} following the methodology of \citet{gordon2020}. The GP uses a custom kernel function based on a power spectrum \citep{anderson1990,kallinger2014} to produce random changes in the granulation coverage.

\begin{figure}
    \centering
    \includegraphics[width=0.5\textwidth]{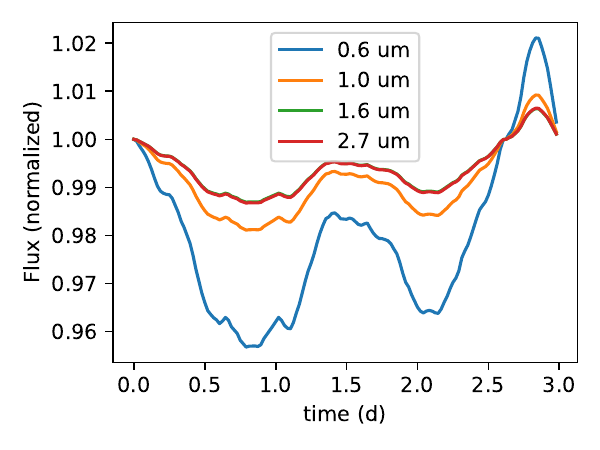}
    \caption{
        Effect of granulation on the light curve of a star. The parameters chosen in this simulation mean that, on average, 10\% of the stellar surface is covered by the cool, inter-granule region with a temperature 300K lower than the surrounding photosphere, and that this coverage varies by about 1\% on a timescale of 6 hours.
        }
    \script{granule_lc.py}
    \label{fig:gran_lc}
\end{figure}

\subsection{Transits \& Limb Darkening}

\begin{figure*}[!htpb]
    \centering
    \includegraphics[width=\textwidth]{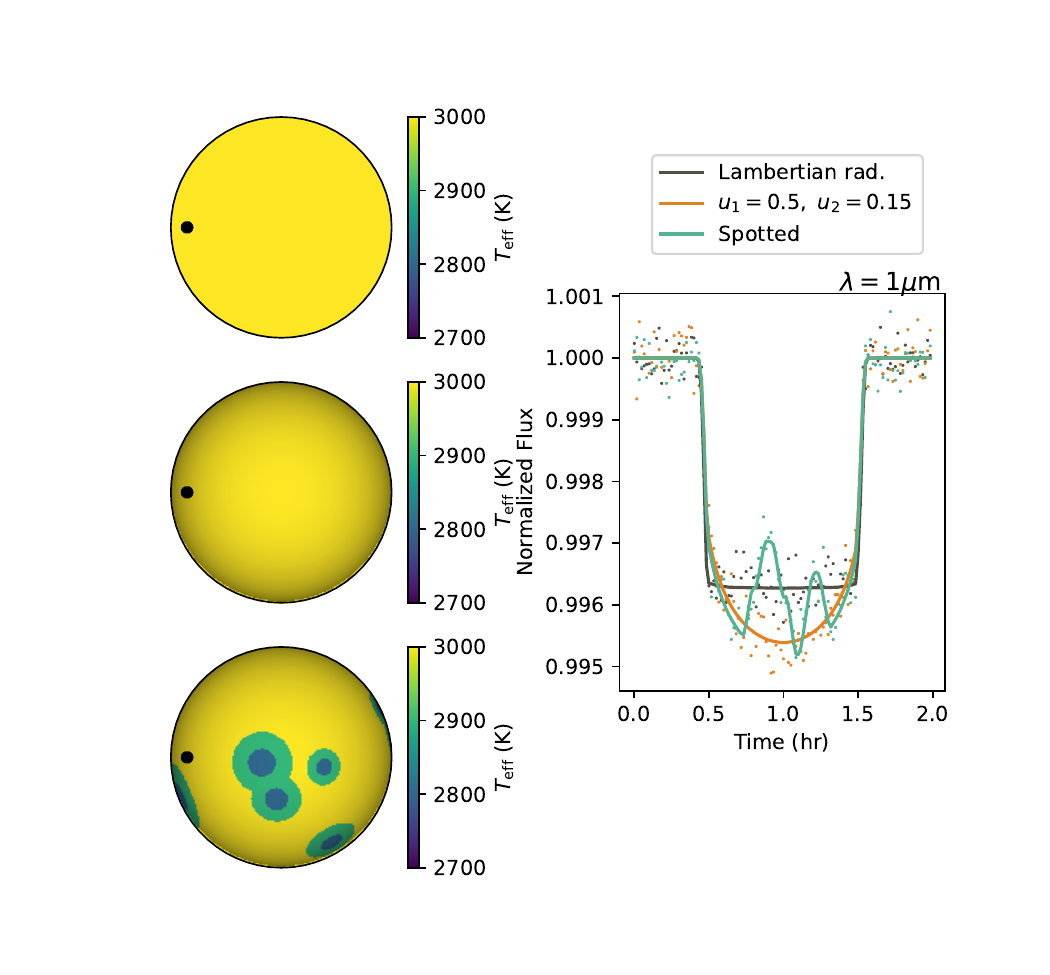}
    \caption{
        {\bf Right}: Transit light curves at 1 ${\rm \mu m}$ for a Lambertian (i.e. no limb darkening) star, one with significant limb darkening \citep[$u_1=0.5,\,u_2=0.15$,][]{espinoza2015}, and one with that same limb darkening law, but with the addition of spots. In each case a 1 $R_e$ planet transits a 0.12 $R_\odot$ ($T_{\rm eff} = 3000$ K) star with an observation cadence of 1 minute. Scattered points show the observation with simulated noise, considering a single wavelength channel from JWST/NIRSpec G395H binned to a resolving power of 270 and a target distance of 5 pc.
        {\bf Left}: Stellar surface maps including translucent masks to show limb darkening and transit. The Lambertian star (top) does not exhibit any limb darkening, and its disk has constant surface brightness. However, the limb-darkened stars (middle and bottom) are much brighter at disk-center than near the limb. The dark circle on the leftmost limb of each star is the region occulted by the planet when it is $0.7^\circ$ from mid-transit. In the spotted star (bottom) the planet crosses over several spots as it moves from left to right. Note in the light curves that while no spot is covered, the transit depth is larger, and the effective radius of the planet appears higher. However, when a spot is crossed, the transit depth decreases dramatically.
    }
    \script{ld_transit.py}
    \label{fig:ld_transit}
\end{figure*}

In the case of a transiting geometry, \texttt{vspec-vsm} is responsible for computing the properties of the portion of the stellar disk that is occulted. It does this by first projecting every visible point (i.e. those for which $\mu > 0$, $\mu$ is the cosine of the angle from disk-center) orthographically onto an $x$-$y$ plane where the stellar disk is represented by the unit circle. The planet, in this projection, is represented by a circle centered at $(a/R_*\,\sin{\phi},~ a/R_*\,\cos{\phi}\,\cos{i})$ with a radius of $R_p/R_*$, for semi-major axis $a$, stellar radius $R_*$, planetary radius $R_p$, phase past midtransit $\phi$, and inclination $i$. It then iterates through each point on the surface, computing the distance in projected coordinates to the center of the occulting disk. Also, for each point, an effective radius is calculated in the projected coordinates
\begin{equation}
    r_{\rm eff} = 2\mu\sqrt{\Omega/(4\pi\,\text{[sr]})}
\end{equation}
where $\Omega$ is the solid angle of each point. Any point for which the distance is less than $R_p/R_* + 2r_{\rm eff}$ from the occulting disk's center is flagged as possibly interacting with the planet. The majority of these points are completely covered by the planet, and those that are less than $R_p/R_* - 2r_{\rm eff}$ from the center of the occultation are immediately marked as such. The remaining points -- those near the edge of the occultation -- are subdivided into a $100\times 100$ point meshgrid. A 2D numerical integral is performed over this grid to compute the fraction of each point that is inside the occulting disk.

 When computing the disk-integrated flux (including flux removed via a transit) the weight of each point is determined by the geometric projected area as well as a quadratic limb darkening law \citep[see][]{espinoza2015}. We treat limb darkening as a weight factor on the contribution of each point to the disk-integrated spectrum that depends on $\mu$. Points that are part of spots and faculae are treated the same as the photospheric points -- their respective weights are increased if they lie near the center and decreased if they lie on the limb. This treatment allows the user to specify two parameters $u_1$ and $u_2$ that describe the flux variation as a function of angle from disk-center. A caveat of this method is that there is no dependence on wavelength. A better treatment would vary the temperature of the stellar model as a function of $\mu$; we simply apply a gray scaling factor to the flux. Figure \ref{fig:ld_transit} compares transit light curves for two different sets of limb darkening parameters.

\section{Stellar Spectral Grids}
\label{sec:gridpolator}

\begin{figure}[h]
    \centering
    \includegraphics[width=0.5\textwidth]{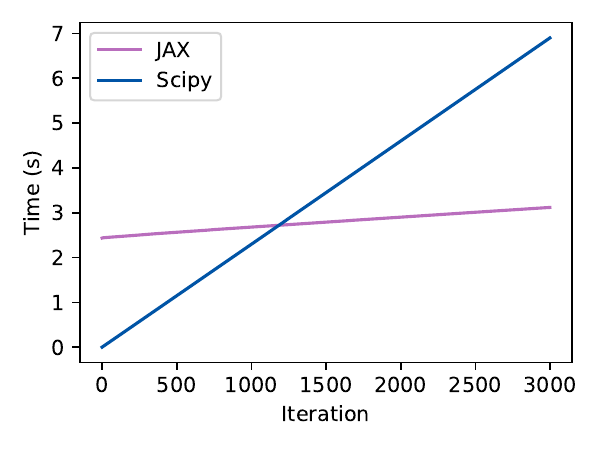}
    \caption{
        Speed comparison between the JAX and SciPy implementations of the GridPolator interpolation backend. Both interpolators use the builtin \vspec{} grid, with \teff ranging from 2800 to 3300K, and wavelengths from 5-12 ${\rm \mu m}$ with $R=100$. SciPy is most efficient for a small number of evaluations, with a constant cost per call. For a large number of evaluations, however, the speed gained via JAX's JIT compilation offsets the cost of performing that compilation.
    }
    \script{jax_v_scipy.py}
    \label{fig:jax_v_scipy}
\end{figure}

\vspec{} utilizes a library of pre-computed stellar models for producing its multi-component composite stellar spectrum; detailed stellar spectral models are computationally expensive to generate, so it is more efficient to interpolate over a discretized grid. This interpolation is handled by the VSPEC-related Python package GridPolator, whose main class \texttt{GridSpectra} acts as a wrapper for the SciPy \citep{virtanen2020} or JAX \citep{bradbury2018} \texttt{RegularGridInterpolator} with a specialized API. This setup accomodates spectral grids of arbitrary dimension, and treats the wavelength axis independently of the other parameters so that each wavelength channel is independent. When a \texttt{Gridspectra} instance's \texttt{evaluate()} method is called, it creates a \texttt{RegularGridInterpolator} instance for each wavelength channel, evaluates that interpolator, and then resamples the returned spectrum at the requested wavelength points. In the case that the JAX implementation is used, the portion of \texttt{evaluate()} that does the interpolation is just-in-time (JIT) compiled to dramatically improve performance on subsequent calls. Figure \ref{fig:jax_v_scipy} compares the computational cost of JAX with that of SciPy.

GridPolator grids can be built arbitrarily by the user, but the package comes with a built-in custom grid of \texttt{PHOENIX} stellar models \citep{allard1994,hauschildt1999,husser2013} created specifically for \vspec{}. Each spectrum spans from 0.1 to 20 microns in steps of $5\times10^{-6}$ microns. The stellar atmospheres used to create the spectra have solar metallicity and a $\log{g}$ of 5.0, with effective temperatures ranging from 2300K to 3900K in steps of 100K. This is the only grid currently supported by \vspec{}, with support for additional grids planned in future releases.
% The second grid is hosted by the Space Telescope Science Institute\footnote{\url{https://archive.stsci.edu/hlsps/reference-atlases/cdbs/grid/phoenix/}}. This grid covers a much larger stellar parameter space, with metallicities ranging from $-4$ to $0.5$, \teff from 2000K to 70\,000K, and $\log{g}$ from 0 to 5.5. These models are also produced by the \texttt{PHOENIX} code. Except for the solar metallicity models, the spectra were computed in 2011 by \citet{allard2003, allard2007} and the solar metallicity models were updated by a new version of \texttt{PHOENIX} in 2021 \citep{allard2012}. \TJ{Look up wavelength ranges}
When the grid is initialized, GridPolator looks for the relevant data files on the user's local system. If they are not present, then the files are downloaded and stored in \texttt{\textasciitilde /.gridpolator/grids/}. 

Within this suite of codes, GridPolator has the added responsibility of binning spectra and producing wavelength axes that match those output by PSG. Wavelength sampling is done using a constant resolving power $R=\lambda/{\Delta \lambda}$. The binning algorithm takes a pair of high-resolution flux and wavelength arrays and resamples the flux to a lower-resolution wavelength axis. PSG specifies the central wavelength of each channel, and so this algorithm defines bin edges to be $\lambda_{\rm cen} \pm {\lambda_{\rm cen}}/{R}$.

\section{Simplified Climate Model}
\label{sec:vspec-gcm}
\vspec{} allows users to run simulations with any climate model they desire (so long as it can be converted into the \texttt{libpypsg.PyGCM} format) and even comes with utilities to work with PSG/GlobES binary files, ExoCAM \citep{wolf2022}, WACCM \citep{marsh2013}, and ExoPlaSim \citep{paradise2022} without any custom code from the user.

However, as a full global circulation model can take days to months to converge, there are many cases where a ``quick and dirty'' 3D climate model may be sufficient. Built into \vspec{} is such a ``faux'' climate model -- one that does not contain any of the circulation physics of the above sophisticated codes, but instead constructs a 3D planetary atmosphere in a fraction of a second from just a few parameters. Such a model can also be advantageous because it is extremely predictable; a feature in the phase curve can easily be traced back to the model parameters that cause it.

\begin{figure}
    \centering
    \includegraphics[width=0.5\textwidth]{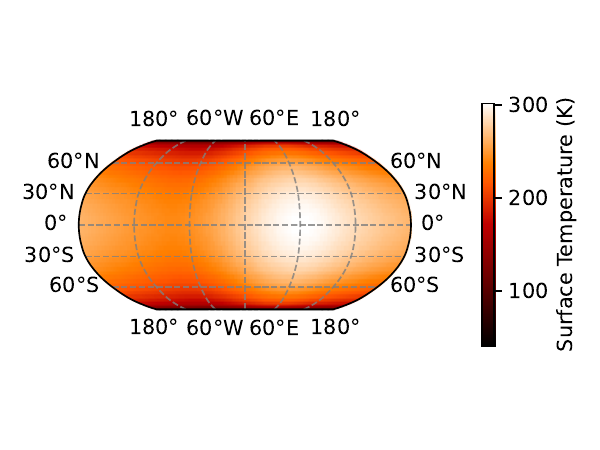}
    \caption{
        Example of a surface temperature map created by the \vspec{} GCM, using values for the solar flux equivalent to Earth, a Bond albedo of 0.3, and $\epsilon = 2\pi$. Notice that the hottest point is offset from the sub-stellar point due to thermal inertia.
    }
    \script{cowan_gcm.py}
    \label{fig:thermal_inertia}
\end{figure}

The \vspec{} GCM centers around a 2D temperature map of the planet's surface. Based on \citet{cowan2011}, this map
is determined exclusively by the incident flux from the star (parameterized by the host \teff~and radius,
and the planet's semimajor axis and Bond albedo) and the unitless thermal redistribution efficiency $\epsilon$.
This efficiency is 0 for a planet that re-radiates instantly and $\gg 1$ for an atmosphere that efficiently redistributes heat to its night side. Figure \ref{fig:thermal_inertia} shows a temperature map example for an Earth-like planet with $\epsilon = 2\pi$. These temperature maps are validated by the energy balance of the planet; \vspec{} computes the incident and emergent flux across the planet and issues a warning if they are not in agreement.

\begin{figure}
    \centering
    \includegraphics[width=0.5\textwidth]{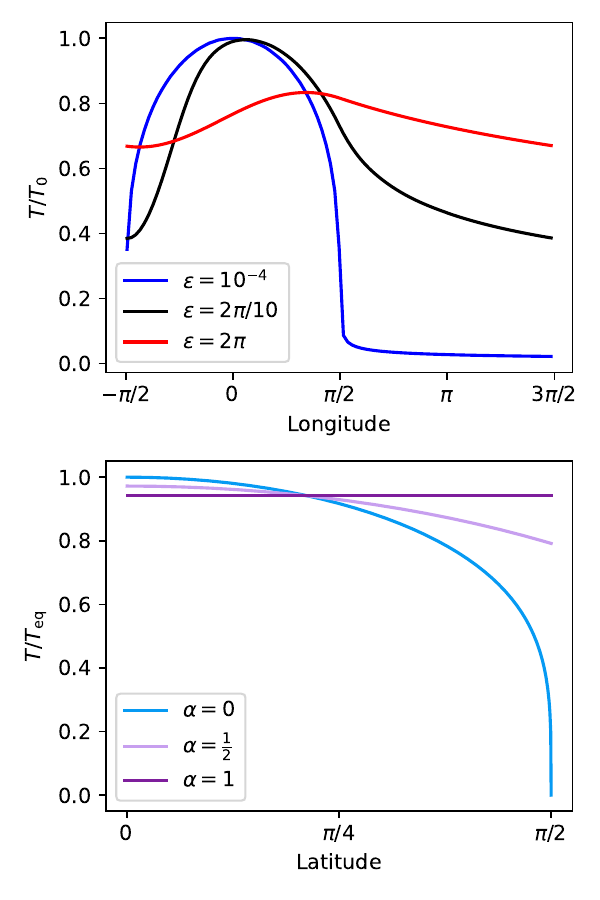}
    \caption{
        {\bf Top}: Reproduction of \citet{cowan2011} Figure 1 using the methods described in the text. These curves show the dimensionless temperature at the equator of a planet as a function of angle from substellar point $\Phi$ for thermal inertia $\epsilon$. Note that the $\epsilon=0$ curve in the original figure is replaced with $\epsilon=10^{-4}$ (see explanation in the text), and we do not include additional curves for mid-latitudes as in the original figure. As $\epsilon$ increases, heat is moved from the dayside onto the nightside and the temperature maximum shifts away from the substellar point.
        {\bf Bottom}: Relative surface temperature as a function of latitude $\lambda$ for varying latitudinal mixing parameter $\alpha$. A value of 1 means the equatorial temperature, as computed in the upper panel. Without mixing ($\alpha=0$), heat is concentrated at the equator. However, as $\alpha$ increases, the equator cools and heat is redistributed towards the poles.
    }
    \script{cowan_fig.py}
    \label{fig:cowan_curves}
\end{figure}

To compute the surface temperature we first compute the temperature at the equator as a function of longitude. This is done by integrating \citet{cowan2011} Equation (10).
\begin{equation}
    \frac{d \tilde{T}}{d \Phi} = \frac{1}{\epsilon}(\max{(\cos{\Phi}, 0)} - \tilde{T}^4)
\end{equation}
Where $\tilde{T}$ is the dimensionless temperature defined as $T/T_0$, or the temperature normalized by a local fiducial value. $\Phi$ is the angle from the substellar point, so we numerically integrate it from $-\pi$ to $\pi$ using one of several different schemes depending on the value of $\epsilon$. For $0 < \epsilon < 1$ we adopt a reflecting initial value problem (IVP) scheme, where we set $\tilde{T}=1$ at $\Phi=\pi$, integrate to $-\pi$, then use the last value as the initial value for an integration from $-\pi$ to $\pi$. Note that for extremely small values of $\epsilon$ (e.g. $10^{-6}$), the integration is slow, and will fail if $\epsilon=0$; we recommended a lower limit of $\epsilon=10^{-4}$. When $1\le \epsilon < 10$ we solve the integral using a boundary value problem (BVP) scheme. For each of these two methods we employ the SciPy \citep{virtanen2020} implementation: \texttt{solve\_ivp()}, employing a 5th order Runga-Kutta scheme \citep{dormand1980} and \texttt{solve\_bvp()}, based on an earlier MATLAB implementation \citep{kierzenka2001}, respectively. When $\epsilon \ge 10$ it is more efficient to use the analytic approximation given by Equation (A2) in \citet{cowan2011}. In Figure \ref{fig:cowan_curves} we show results for several different values of $\epsilon$.

\subsection{Extending to 3 Dimensions}

The treatment by \citet{cowan2011} includes no mixing in the latitudinal direction, nor a vertical temperature profile. To allow for variations in pressure and inclination, we augment their treatment with several simple approximations for latitudinal and vertical temperature variations. However, it should be noted that these assumptions produce a discontinuity at the poles, which is non-physical. A more physically-motivated model of latitudinal mixing (e.g., including a diffusion parameter in the heat transport equation) would avoid this caveat, but would also be more expensive to produce.

We add a parameter $\alpha$ that is defined by the ratio between the polar temperature and the equatorial temperature at any given longitude. We define $f(\lambda;\alpha)$ to be the temperature as a function of latitude $\lambda$ at an arbitrary longitude; therefore if the surface is in equilibrium, the energy balance requires the outgoing energy flux to be independent of $\alpha$:
\begin{equation}
    \label{eq:e_out}
    E_{\rm out} = \sigma \int_0^{\pi/2} f(\lambda; \alpha)^4 \cos{\lambda}\,d\lambda
\end{equation}
where $E_{\rm out}$ is the outbound energy flux of an infinitesimal sliver of differential longitude. In the case of $\alpha=0$ (i.e. no mixing) we can treat each latitude value independently, so the equilibrium temperature is
\begin{equation}
    f(\lambda; 0) = f(0;0) \cos^{1/4}{\lambda}
\end{equation}
where $f(0;0)$ is the previously computed temperature at the equator (see Figure \ref{fig:cowan_curves}). Plugging this into Equation (\ref{eq:e_out}), we see
\begin{equation}
    \label{eq:nomix}
    \begin{array}{cl}
        E_{\rm out} & = \sigma \int_0^{\pi/2} (f(0;0) \cos^{1/4}{\lambda})^4\cos{\lambda}\,d\lambda \\
        & = \sigma f^4(0;0) \int_0^{\pi/2} \cos^{2}{\lambda}\,d\lambda \\
        & = \frac{\pi}{4}\sigma f^4(0;0)
    \end{array}
\end{equation}

If instead we let $f(\lambda;1)$ be a constant function where the equator and pole have equal temperatures, then
\begin{equation}
    \label{eq:allmix}
    \begin{array}{cl}
        E_{\rm out} & = \sigma \int_0^{\pi/2} f(\lambda;1)^4 \cos{\lambda}\,d\lambda \\
        & = \sigma f^4(0;1) \int_0^{\pi/2} \cos{\lambda}\,d\lambda \\
        & = \sigma f^4(0;1)
    \end{array}
\end{equation}

We calculate $f(0,1)$ by setting Equations (\ref{eq:nomix} \& \ref{eq:allmix}) equal to find
\begin{equation}
    f(0;1) = \left(\frac{\pi}{4}\right)^{1/4} f(0;0) \approx 0.94 \,f(0;0)
\end{equation}

We can now write a general parameterization for the temperature as a function of latitude
\begin{equation}
    \label{eq:temp_gen}
    f(\lambda;\alpha) = f(0;0)\,\left(\alpha \frac{\pi}{4} + (1-\alpha) \cos{\lambda}\right)^{1/4}
\end{equation}

%\subsection{Pressure and Temperature}
Combining Equation (\ref{eq:temp_gen}) with the numerically-computed equatorial temperatures yields a 2D temperature map that covers the surface of the planet. We construct a 3D atmosphere on top of this surface assuming the atmosphere is adiabatic with index $\gamma$. The user supplies both the adiabatic index and the surface pressure $P_{\rm surf}$. The temperature at some altitude $z$ is given by
\begin{equation}
    \label{eq:tz}
    T_z = T_{\rm surf}\,\left( \frac{P_z}{P_{\rm surf}} \right)^{1-\frac{1}{\gamma}}
\end{equation}

Prescribing the temperature profile via Equation (\ref{eq:tz}) is, of course, an oversimplification. The profile does not consider the energy transport between layers of the atmosphere, and will violate the energy balance of the planet. We emphasize that this model is meant to approximate an atmosphere quickly, not be a self-consistent climate model.

Finally, we populate our atmosphere with molecules. The user specifies the species and abundance, and the atmosphere is filled accordingly with a constant molecular suite.

\section{Examples \label{sec:examples}}

\subsection{Transit of a spotted star}

\begin{figure}
    \centering
    \includegraphics[width=0.5\textwidth]{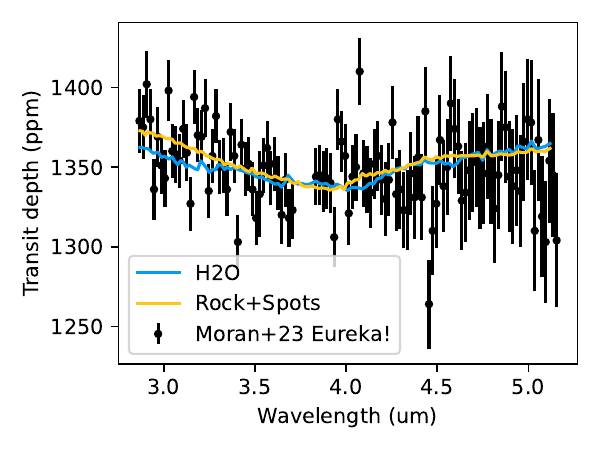}
    \caption{
        The results from our spotted transit experiment. We find that the addition of spots to a star's surface can produce spectral features in what would otherwise be a flat transit spectrum due to a lack of atmosphere or other mitigating issues such as planetary aerosols.  These features mimic those that would be produced due to molecular absorption in a planetary atmosphere. This is in agreement with the results of \citet{moran2023}.
        }
    \script{moran23.py}
    \label{fig:moran_transit}
\end{figure}

In this section we will simulate the transit of a atmosphere-less occulting disk across a spotted star to demonstrate how an inhomogeneous
stellar surface can lead to a false atmospheric signal -- a phenomenon known as the ``transit light source effect'' \citep{rackham2018}.

During primary transit of a planet with an atmosphere, the atmosphere is illuminated by the portion of the star directly behind the planet. Ideally, the depth of the transit should be measured with respect to the spectrum of the region of the stellar surface covered by the planet. However, the disk-integrated stellar spectrum (as measured before an after the transit) is used as the baseline. In the case that the stellar surface is not homogeneous, the transit signal is modulated by the uncorrected stellar inhomogeneities. In this example, we will reproduce the TLS results by \citet{moran2023}, who contended with this degeneracy with JWST/NIRSpec observations during a transit of super-Earth GJ 486b.

Whenever possible, we draw parameter values from \citet{moran2023} and the NExScI rchive\footnote{\url{https://exoplanetarchive.ipac.caltech.edu/overview/GJ486b}}. We use the JWST NIRSpec/G395H instrument setup from PSG with an observing time of 3.53 hours with an 8 minute cadence to match \citet{moran2023}, but set $R=200$ to account for the resolution of their reduction. For simplicity, we do not include limb darkening or stellar rotation.

We performed two \vspec{} simulations based on the two scenarios described in \citet{moran2023}:
\begin{enumerate}
    \item 1 bar \ce{H2O} isothermal atmosphere, no spots. \label{case:1barh2o}
    \item No atmosphere, 2.5\% coverage by 2700 K spots \label{case:no_atmo}
\end{enumerate}

In Case \ref{case:1barh2o} we expect to see an absorption spectrum characteristic of \ce{H2O}. Case \ref{case:no_atmo} is more complicated, though in the absence of spots we would expect a flat spectrum. Note also that we disabled \ce{H2O}-\ce{H2O} collision-induced absorption (CIA) for
these simulations because the effect was not considered in the original analysis of GJ 486b. This is a simplified and highly controlled experiment, so there is no need to use standard transit light curve fitting techniques. We are able to compute the transit depth as $(f_\text{pre}-f_\text{tr})/f_\text{pre}$, where $f_\text{pre}$ is the total spectrum before the transit, at an epoch chosen by visual inspection, and $f_\text{tr}$ is the total spectrum during the transit, again chosen by visual inspection. A light curve fitting code would be needed if we had introduced stellar rotation or limb darkening.

We find that, as expected, the addition of spots changes the transit depth in a way that depends on wavelength -- mimicking light that would be lost to absorbers in a planetary atmosphere. Figure \ref{fig:moran_transit} shows us that, in this case, we observe false absorption signals; this is due to a temporary increase in the spotted fraction of the surface relative to the quiet photosphere. However, if a spot was blocked by the planet we might see similar features inverted to mimic emission.

\subsection{Phase Curve}
\begin{figure}[!htbp]
    \centering
    \includegraphics[width=0.5\textwidth]{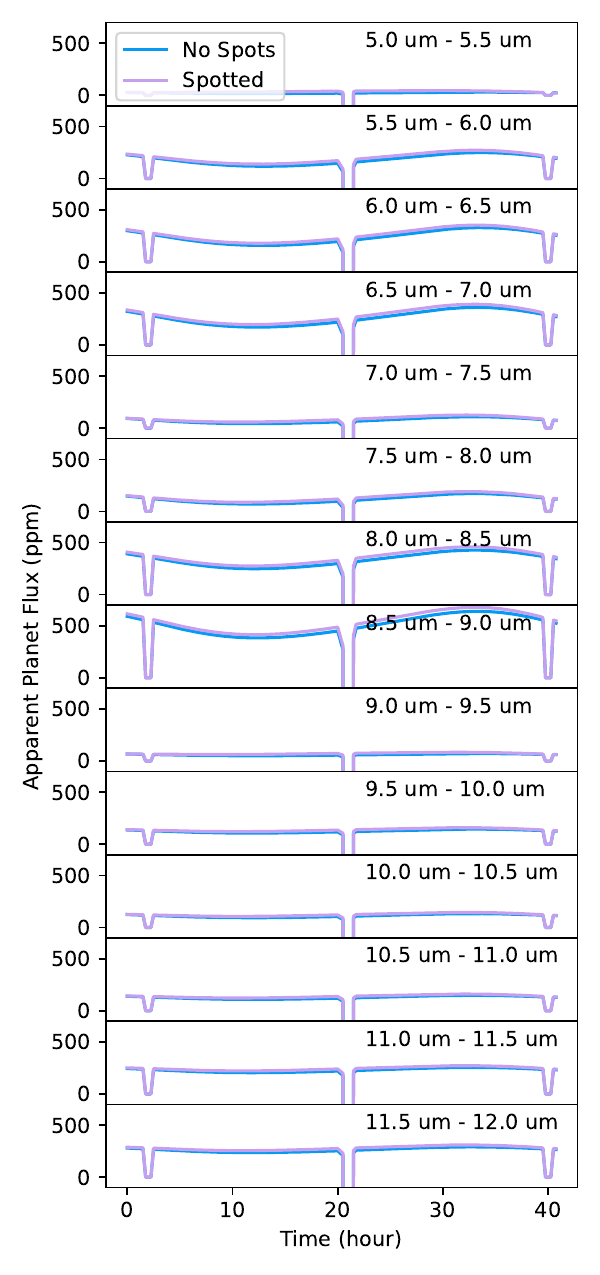}
    \caption{
        A simulation of a MIRI phase curve for GJ 1214b, binned by wavelength; compare to real data published by \citet{kempton2023}(Extended Data Figure 1). While the addition of stellar variability does have a visible effect on the light curve, it is much smaller than the thermal emission produced by the exoplanet.
        }
    \script{kempton23.py}
    \label{fig:gj1214b}
\end{figure}

Analysis of JWST MIRI-LRS phase curves of GJ 1214b by \citet{kempton2023} did not consider stellar contamination because of the slow rotation rate of the host star \citep[approximately 1/80$^{\text{th}}$ the orbital frequency,][]{cloutier2021}. In this example, we simulate these observations using \vspec{} and determine whether this is a reasonable assumption.

We initialized \vspec{} to simulate an observation of the system using the same instrument setup and observation parameters as \citet{kempton2023}. We used stellar and planetary parameters from the NExScI Exoplanet Archive\footnote{\url{https://exoplanetarchive.ipac.caltech.edu/overview/GJ1214b}}. The VSPEC GCM (see Section \ref{sec:vspec-gcm}) used has a thermal inertia of $\epsilon = 6$ and a 1 bar \ce{CO2} atmosphere, appropriate for a high-metallicity \ce{CO2}-rich atmosphere scenario.

We initialized two stellar models: one with nothing more than a 3250 K surface, and another with 20\% coverage by 2700 K spots. After running the \vspec{} simulations, we binned the simulated data into 0.5 $\mu m$ wavelength channels and normalized each channel to the two eclipses, similar to standard data analysis techniques; this assumes that the star behaves linearly between eclipse measurements and therefore attributes any deviation to the planet.

We show in Figure \ref{fig:gj1214b} that the stellar properties assumed in this example do not generate enough contamination to
significantly change the observed phase curve of GJ 1214b. At worst (i.e. near transit), the observed spectrum varies by 50 ppm due to stellar
contamination -- compared to the 700 ppm relative flux of the planet at $8.5 {\mu m}$. 

Similar studies could be done to assess the potential for
stellar contamination in other exoplanetary systems in order to judge fitness as a target; in particular, faster stellar rotation should produce much larger contamination signals that may approach or dwarf the planetary signal.

\section{Conclusion}
\label{sec:conclusion}
We present a suite of codes to model observations of exoplanets with 3D atmospheres orbiting variable stars. The main package, \vspec{}, produces combined-light phase curves by interfacing with an ensemble of support packages whose responsibilities range from grid interpolation to interfacing with PSG. These codes are all open-source and available to the public on GitHub and the Python Package Index (PyPI).

While all of these codes are stable, development is ongoing. Future work on \vspec{} will focus on improving performance so that it can be used to produce phase-curve grids for retrievals. In general, a phase curve retrieval will not require the stellar portion of VSPEC's output, and if there is no need for both thermal and reflected spectra, then only one PSG calculation is required per epoch. This could reduce \vspec{}'s computational cost by approximately a factor of 4 purely by skipping unnecessary steps.

A future goal of \texttt{vspec-vsm} development should be to create a common API for externally-developed physics-based variable star surface models, so that \vspec{} can interface with those models as well. GridPolator is currently in the process of adding support for more built-in stellar grids, which are more complex and cover a larger range of spectral types than the current \vspec{} grid. However, as the size of these grids grows, it becomes less efficient to build the grid prior to each run. Future versions will support caching of grids locally using standard Python object serialization libraries such as \texttt{pickle}\footnote{\url{https://docs.python.org/3/library/pickle.html}}. This method would allow interpolator-oriented Python objects to be recalled nearly instantly (limited by hard drive read-write speeds) rather than the expensive process of re-initialization.

\texttt{libpypsg} has the most to offer the exoplanet community as a stand-alone package, but currently lacks support for much of the data that PSG is capable of producing. Its development so far has focused around the \vspec{} use case, so future work will center around making it more useful to the majority of PSG users.

\section{Acknowledgements}
The authors would like to thank Peter Hauschildt for generating a custom grid of PHOENIX stellar models.

T.J. acknowledges support from the Exoplanet Modeling and Analysis Center \citep[EMAC, ][]{renaud2022}.

The authors would like to thank the Sellers Exoplanet Environments Collaboration (SEEC) and Exoplanet Spectroscopy Technologies (ExoSpec) ISFM teams at NASA's Goddard Space Flight Center for their consistent financial and collaborative support.

This manuscript was typeset using the open-science software \showyourwork, which makes the entire manuscript, including each figure, completely reproducible by anyone with access to the git repository at \url{https://github.com/VSPEC-Collab/vspec-paper}.

This work made use of Astropy:\footnote{\url{http://www.astropy.org}} a community-developed core Python package and an ecosystem of tools and resources for astronomy \citep{astropycollaboration2013,astropycollaboration2018,astropycollaboration2022}.

The figures in this article were created using the Matplotlib \citep{hunter2007}, NumPy \citep{harris2020}, and Cartopy \citep{metoffice2010} libraries. Figure \ref{fig:vspec-diagram} was made using Lucidchart\footnote{\url{https://www.lucidchart.com}}. The top panel of Figure \ref{fig:fac_struct} was created using the open-source vector graphics software Inkscape\footnote{\url{https://inkscape.org/}}.

\section{Author Contributions} % see https://www.elsevier.com/researcher/author/policies-and-guidelines/credit-author-statement
{\bf Ted M Johnson:} Conceptualization, Methodology, Software, Writing - Original Draft
{\bf Cameron Kelahan:} Conceptualization, Methodology, Software, Writing - Review \& Editing
{\bf Avi M. Mandell:} Conceptualization, Methodology, Writing - Original Draft, Writing - Review \& Editing, Supervision, Project administration, Funding acquisition
{\bf Ashraf Dhahbi:} Software
{\bf Tobi Hammond:} Software
{\bf Thomas Barclay:} Methodology, Software
{\bf Veselin B. Kostov:} Methodology
{\bf Geronimo L. Villanueva:} Resources, Writing - Review \& Editing, Supervision

\bibliographystyle{elsarticle-harv}
\bibliography{syw,VSPEC}

\end{document}